\documentclass{kapproc} 

\kluwerbib

\begin{document}

\articletitle{Emission Processes of High Energy Gamma Rays from
Gamma-Ray Bursts}


\author{X. Y. Wang, Z. G. Dai and T. Lu}
\affil{Department of Astronomy, Nanjing University, Nanjing
210093, China} \email{xywang@nju.edu.cn;
daizigao@public1.ptt.js.cn; tlu@nju.edu.cn}


\begin{abstract}
Synchrotron self-Compton (SSC) process in the reverse shocks of
gamma-ray bursts is suggested to be responsible for the observed
prompt high-energy gamma-ray emissions from several gamma-ray
bursts. We find that the SSC emission from the reverse shocks
dominates over other emission processes in  energy bands from tens
of MeV to tens of GeV, for a wide range of shock parameters. This
model is favorable for escape of energetic photons from the
emitting regions due to a lower internal pair-production optical
depth, as the characteristic size of the reverse shock region is
much larger than that of internal shocks. We predict that, in this
model, the prompt high-energy emissions are correlated with the
prompt optical flashes, which can be test in the forthcoming GLAST
era.

\end{abstract}


\section {Introduction}
 EGRET has detected  prompt emission above $30{\rm MeV}$
 from several  bright GRBs
triggered by BATSE (Catelli et al. 1998), among which GeV photons
have been detected from GRB930131 (Sommer et al. 1994; Ryan et al.
1994) and GRB940217 (Hurley et al. 1994).  GRB940217 even exhibits
delayed GeV emission 90 minutes after the trigger. Two classes of
models have been proposed to explain the delayed and prompt GeV
emissions.

One is related to "Hadron processes". It is widely assumed that
GRB shocks (internal and/or external) can accelerate protons  to
very high energies, a proposed mechanism for the production of
ultra-high energy cosmic rays (Waxman 1995; Vietri 1995). The
photo-meson processes (Waxman \& Bahcall 1997; B${\rm \ddot
o}$ttcher \& Dermer 1998) or synchrotron radiations of the protons
(Vietri 1997; Totani 1998a,b) have been suggested to be
responsible for the GeV emissions. Katz (1994) even suggested that
the impact of the fireball on a dense clouds could produce
high-energy gamma-ray emission via $\pi^0$ decay process.

Another class is related to the inverse Compton processes in GRB
shocks, including internal shocks and external shocks. In the
standard picture of GRBs, sub-MeV gamma-rays are formed in
internal shocks. When the relativistic ejecta encounters the
external medium,
 a relativistic { forward} shock expands into the external medium and a { reverse}
 shock moves into and heats the fireball ejecta. The forward shock continuously heats fresh gas and accelerates
electrons, producing long-term afterglows through the synchrotron
emission (e.g. van Paradijs et al. 2000). A strong prompt optical
flash (Akerlof et al. 1999) and late time radio flare behavior
(Kulkarni et al. 1999), accompanying GRB990123, have been
attributed to the synchrotron emissions  from the reverse shock
(Sari \& Piran 1999; {M\'esz\'aros} \& Rees 1999). Papathanassiou
\& Meszaros (1994) proposed that electron IC processes in internal
shocks produce GeV emissions, while Meszaros, Rees \&
Papathanassiou (1994), Dermer et al. (2000) and  Zhang \& Meszaros
(2001) suggest that electron IC processes in forward shocks may be
responsible for the prompt and delayed GeV emissions.

We here suggest an alternative mechanism,  that is the SSC
emissions from reverse shocks. AS shown below, the SSC emission
from the reverse shocks dominates over other emission processes in
energy bands from tens of MeV to tens of GeV, for a wide range of
shock parameters (Wang, Dai \& Lu 2001a,b). Moreover, it involves
a much larger emitting size, hence a lower internal
pair-production optical depth than models related to internal
shocks.

In section 2, we consider the attenuation effects of high energy
photons in internal and external shocks. We  analytically study
(section 3) the high energy $\gamma$-ray emission from the SSC
process in reverse shocks and in section 4 numerically calculate
the SSC radiation components in the reverse shocks and compare it
with the SSC emission from forward shocks and  another two
combined-IC processes ( M\'esz\'aros, Rees \& Papathanassiou
1994), i.e. scatterings of reverse shock photons on the forward
shocked electrons and forward shock photons on the reversely
shocked electrons.

\section{Attenuation of high energy photons in internal and
external shocks}

It has been suggested that high energy photons are produced in
internal shock regions (e.g. Papathanassiou \& Meszaros 1994;
Waxman \& Bahcall 1997). However, as we show below, energetic GeV
photons  may suffer strong attenuation by other photons  produced
in internal shocks. The observed photon spectra of GRBs can be
approximated as power law, with a high-energy part of the form
$dn/d\varepsilon\propto \varepsilon^{-\beta}$ for
$\varepsilon>\varepsilon_b$, where $\varepsilon=h\nu/m_e c^2$ is
the photon energy in units of the electron's rest energy and
$\varepsilon_b\sim1$ is the break energy above which $\beta\sim
2-3$. As a photon with energy $\varepsilon$ will annihilate any
photon with energy above $\varepsilon_c=\eta^2/\varepsilon$ (Where
$\eta$ is the Lorentz factor of GRB fireball), its optical depth
is given by $\tau_{\gamma\gamma}={(11/180)\sigma_T N_c}/{4\pi
R_{in}^2}$ (Lithwick \& Sari 2001), where
$N_c\equiv\int^{\infty}_{\varepsilon_c}dn/d\varepsilon
d\varepsilon$ and $R_{in}=2\eta^2 c \delta t$ is the radius of
internal shocks, with $\delta t$ being the rapid variability
timescale in GRB light curves. Setting
$\tau_{\gamma\gamma}(\varepsilon_c)=1$, we derive the cut-off
energy
\begin{equation}
h\nu_c=0.5{\rm GeV}\delta
t_{-1}^{\frac{2}{\beta-1}}E_{\gamma,53}^{-\frac{1}{\beta-1}}\varepsilon_b^{-\frac{\beta-2}
{\beta-1}}\eta_{300}^{\frac{2\beta+2}{\beta+1}}
\end{equation}
where $E_{\gamma}\equiv 10^{53}{\,\rm erg\,} E_{\gamma,53}=\int
\varepsilon m_e c^2N_c d\varepsilon$ is the burst energy in
gamma-rays, $\delta t\equiv 10^{-1} {\, \rm s\,} \delta t_{-1}$,
$\eta\equiv 300\,\eta_{300}$ and the numerical coefficient on the
right hand side corresponds to $\beta=2.2$.

However, this problem will find a natural solution if the
high-energy photons are formed in external shocks because the size
of the latter is 2-3 orders of magnitude larger than that of
internal shocks.

\section{The analytic estimate}
The synchrotron emission spectrum at the deceleration time
($t_{dec}$) can be described by two break frequencies and the peak
flux (for  redshift $z=1$):
\begin{equation}
\nu_{m}^{rs}=\frac{\Gamma({\gamma_m^{rs}})^2{eB'}}{2\pi{m_e}c}=
6.4\times10^{15}{\rm Hz\,}{\xi_{e,0.6}^2}{\xi_{B,-2}^{1/2}}
\eta_{300}^2{n_0^{1/2}},
\end{equation}
\begin{equation}
\nu_c^{rs}=\frac{10^{17}\rm Hz}{(Y_{rs}+1)^2}{\xi_{B,-2}^{-3/2}}
\eta_{300}^{-4}{n_0^{-3/2}}(\frac{t_{dec}}{10\rm sec})^{-2}
\end{equation}
\begin{equation}
f_m^{rs}=1.5{\rm Jy \,
}h_{65}^2{\xi_{B,-2}^{1/2}}{\eta_{300}^{-1}}{n_0^{-1/4}}{E_{53}^{5/4}}(\frac
{{t_{dec}}}{10{\rm sec}})^{-3/4},
\end{equation}
where $E=10^{53}E_{53}{\rm erg}$ is the shock isotropic energy,
$n=1n_0{\rm cm^{-3}}$ is the number density of the interstellar
medium, $z$ is the redshift of the GRB source, $B'=12{\rm
G}{\xi_B}_{0.01}^{1/2}\eta_{300}n_0^{1/2}$, is the magnetic field
in the comving frame, $\xi_e\equiv0.6\xi_{e,0.6}$ and $\xi_B\equiv
0.01\xi_{B,-2}$ are the equipartition values of electron and
magnetic energies respectively, and a flat universe with zero
cosmological constant and $H_0=65h_{65} {\rm Km~s^{-1}{Mpc}^{-1}}$
is assumed. The Compton parameter Y, expressing the cooling rate
of electrons due to inverse Compton effect, is defined as
$Y=\frac{4}{3}\tau_{e}\int{\gamma}^2 \bar{N}_e(\gamma)d{\gamma}$,
where$\bar{N}_e(\gamma)$ is the normalized electron distribution
and $\tau_{e}$ is the optical thickness to electron scattering.

Then we derive the SSC spectrum of the reverse shocks:
\begin{equation}
\nu_m^{rs,IC}=2(\gamma_m^{rs})^2{\nu_m^{rs}}=1.0\times10^{21}{\rm
Hz}\xi_{e,0.6}^4\xi_{B,-2}^{1/2}{\eta_{300}^2}{n_0^{1/2}}
{\bar\gamma_{rs}^2},
\end{equation}
\begin{equation}
\nu_c^{rs,IC}=2(\gamma_c^{rs})^2\nu_c^{rs}=2.1\times10^{22}{\rm
Hz}
\xi_{B,-2}^{-7/2}{\eta_{300}^{2/3}}{n_0^{-13/6}}{E_{53}^{-4/3}},
\end{equation}
\begin{equation}
f_{max}^{rs,IC}=\tau_e^{rs}f_{m}^{rs}=2.6\times10^{-8}{\rm
erg~cm^{-2}s^{-1}MeV^{-1}}h_{65}^2
E_{53}^{4/3}{n_0^{7/6}}{\eta_{300}^{4/3}}\xi_{B,-2}^{1/2}.
\end{equation}
For typical parameters $\xi_e=0.6$, $\xi_B=0.01$, $p=2.5$ and
$n=1$, we give the flux of the inverse Compton component at two
representative frequencies:
\begin{eqnarray}
f^{rs,IC}(\varepsilon=100{\rm MeV})=1.0\times10^{-9}{\rm
erg~cm^{-2}s^{-1}MeV^{-1}}
E_{53}^{4/3};  \nonumber \\
f^{rs,IC}(\varepsilon=1{\rm GeV})=1.5\times10^{-10}{\rm
erg~cm^{-2}s^{-1}MeV^{-1}} E_{53}^{2/3}.
\end{eqnarray}
As a comparison, the derived high energy flux of the synchrotron
and SSC emissions from  forward shocks are, respectively,
\begin{eqnarray}
f^{fs}(\varepsilon=100{\rm MeV})=1.0\times10^{-10}{\rm
erg~cm^{-2}s^{-1}MeV^{-1}}
E_{53};  \nonumber \\
f^{fs}(\varepsilon=1{\rm GeV})=0.5\times10^{-11}{\rm
erg~cm^{-2}s^{-1}MeV^{-1}} E_{53},
\end{eqnarray}
\begin{equation}
f_m^{fs,IC}=3\times10^{-13} {\rm erg~cm^{-2}s^{-1}MeV^{-1}}.
\end{equation}
Therefore, we conclude that for the typical parameter values of
the shock and the surrounding medium, the synchrotron self-Compton
emission from the reverse shock dominates over the synchrotron and
synchrotron self-Compton emissions from the forward shock at  high
energy gamma-ray bands. Our result is different from that of
Dermer et al. (2000), who argue that the synchrotron self-Compton
emission from the forward shock may be responsible for the prompt
and delayed high energy gamma-ray emission.
 The key point of the difference is that
they considered a rather dense circumburst medium with number
density
 $n\sim100{\rm cm^{-3}}$, while
we consider a typical interstellar medium  with $n\sim1{\rm
cm^{-3}}$.

As an example, we try to fit the high-energy emissions from
GRB930131. The  photon spectrum of GRB990131 can be described by
${dn}/{d\varepsilon}\sim7.4\times10^{-6}{\rm
photons~(cm~s~MeV)^{-1}} (\varepsilon/147{\rm
MeV})^{-2.07\pm0.36}$ (Sommer et al. 1994), while our model
prediction for $\varepsilon>h{\nu_c^{rs,IC}}$ is
${dn}/{d\varepsilon}\sim 2.2\times10^{-6}{\rm
photons~(cm~s~MeV)^{-1}}(\varepsilon/147{\rm MeV})^{-2.25}
E_{53}^{2/3}$. Thus, if the fireball shock energy
$E\sim4\times10^{53}{\rm erg}$ and other parameters such as
$\xi_e$, $\xi_B$, $\eta$, $z$ and the number density $n$ of the
surrounding medium take the above representative values, then both
the flux level and the spectrum agree well with the observations.

\section{Numerical Result}
Four IC processes, including the synchrotron self-Compton (SSC)
 processes in GRB forward
and reverse shocks, and two combined-IC processes (i.e. scattering
of reverse shock photons on the electrons in  forward shocks and
forward shock photons on the electrons in reverse shocks), are
considered now (Wang, Dai \& Lu 2001b).

For single scattering, the IC volume emissivity in the comoving
frame for a distribution $N(\gamma) $ of scattering electrons is
given by (Rybicki \& Lightman 1979; Sari \& Esin 2001)
\begin{equation}
j_{\nu'}^{'IC}=3\sigma_T\int_{\gamma_{min}}^{\gamma_{max}}d\gamma{N(\gamma)}
\int_0^1{dx}{g(x)}\bar{f}'_{\nu'_s}(x),
\end{equation}
where $x\equiv\nu'/4\gamma^2{\nu'_s}$, $\bar{f}'_{\nu'_s}$ is the
incident specific flux at the shock front in the comoving frame,
and $g(x)=1+x+2x{\rm ln}(x)-2x^2$ reflects the angular dependence
of the scattering cross section for $\gamma_e\gg1$ (Blumenthal \&
Gould 1970). Noting that
$f_{\nu'}^{'IC}=j_{\nu'}^{'IC}4\pi{r^2}\Delta{r'}/4\pi{D^2}$ and
the synchrotron flux
$f'_{\nu'}=\bar{f}'_{\nu'_s}4\pi{r^2}/4\pi{D^2}$, where
$\Delta{r'}$ is the comoving width of the shocked shell or ISM
medium and $D$ is source distance, we obtain the IC flux in the
observer frame
\begin{equation}
f_{\nu}^{IC}=3\Delta{r'}\sigma_T\int_{\gamma_{min}}^{\gamma_{max}}d\gamma{N(\gamma)}
\int_0^1{dx}{g(x)}f_{\nu}(x)
\end{equation}
by transforming Eq.(11) into the observer frame.

Apart from the SSC scattering processes in the reverse and forward
shocks, another two combined-IC scattering processes
 are also present.
 Because approximately one-half of the photons
 arised in one shock region will diffuse into the another shock
 region from the point of view of the comoving frame, the IC flux Eq.(12) for
 the combined-IC scatterings
  should be
 divided by a factor of two.  Though  the scattered photons
 move isotropically in the comoving frame, the beaming effect makes these photons
 moving along the direction to the observer.

Our main calculation results are as follows:

i) The IC spectral flux from a relativistic shell expanding into
an ISM at the deceleration time are shown in Fig. 1. Typical shock
parameters are used: $E=10^{53}{\rm erg}$, { $\xi_e=0.6$},
$\xi_B=0.01$, $p=2.5$ and $n=1$. It can be clearly seen that the
SSC from the reverse shock dominates over the other three IC
components at gamma-ray bands less than  a few tens of GeV with {
a peak around a few {\rm MeV}}.

\begin{figure}
\vskip 1.5in \includegraphics{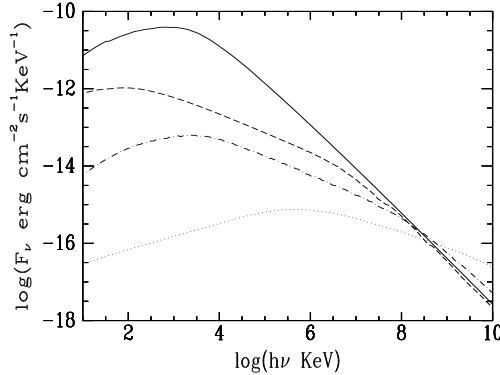}
 \vskip 1.5cm
 \caption{The spectra of the IC emissions at the reverse shock peak time
for typical shock parameters. The {\em solid } and {\em dotted}
curves represent the SSC emissions from the reverse shock and
forward shock, respectively. Also plotted are the IC emissions of
scatterings of reverse shock photons on the forward shock
electrons ({\em dash-dotted curve}) and forward shock photons on
the reversely shocked electrons ({\em dashed curve}).}
\end{figure}
\noindent



 ii) In Fig.2, we present the energy spectra
($\nu{f_{\nu}^{IC}}$) of the IC emissions with various shock
parameters. We find that a) for a wide range of shock parameters,
the SSC component from reverse shocks is the most important at
energy bands from tens of MeV to tens of GeV, to which EGRET is
sensitive. b) For small value of $p$ (e.g. $p=2.2$), the SSC
emission from the reverse shock dominates over the synchrotron and
IC processes even in the TeV energy bands (see Fig. 2(d)). Fig.2
also suggest that strong TeV emission should also be emitted from
the two combined-IC  and forward shock SSC  processes for most
GRBs. For a moderate steep distribution of the shocked electrons
(e.g. $p=2.5$), the combined-IC and/or forward shock SSC become
increasing dominated at TeV bands.
 However, it would only be detected from nearby, low-redshift
bursts for which the attenuation due to intergalactic infrared
emission is small.
\begin{figure}
\vskip 1.5in \includegraphics{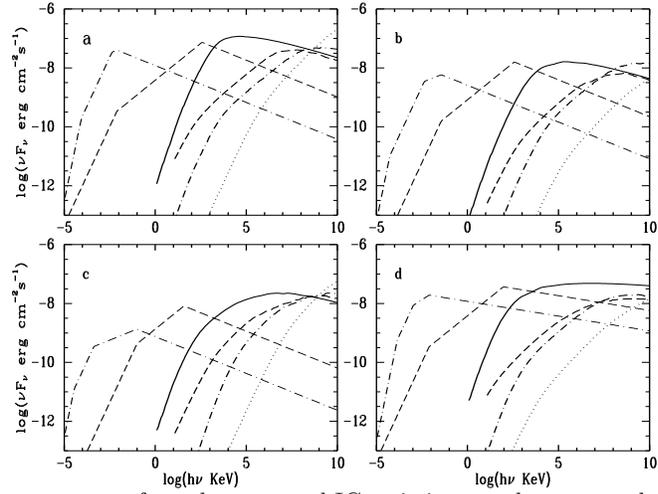}
 \vskip 3.0cm

 \caption{The energy spectra of  synchrotron and IC emissions
  at the reverse shock peak time for the ISM circumburst environment case with various
  shock parameters: a)$E=10^{53}{\, \rm erg}$, $\xi_e=0.6$, $\xi_B=0.01$, $p=2.5$
  and $n_1=1$; b)$E=10^{52}{\, \rm erg}$, $\xi_e=0.6$, $\xi_B=0.01$, $p=2.5$
  and $n_1=1$;
  c)$E=10^{53}{\, \rm erg}$, $\xi_e=0.6$, $\xi_B=10^{-4}$, $p=2.5$ and $n_1=1$;
  d)$E=10^{53}{\, \rm erg}$, $\xi_e=0.6$, $\xi_B=0.01$, $p=2.2$ and $n_1=1$.
  The   {\em thin dash-dotted} and {\em dashed curves } represent the synchrotron
  spectra of the reverse shock and forward shock, respectively. The four IC spectra
  are shown by the curves in the same way as in Fig. 1.}
\end{figure}
\noindent
\begin{figure}
\vskip 1.5in \includegraphics{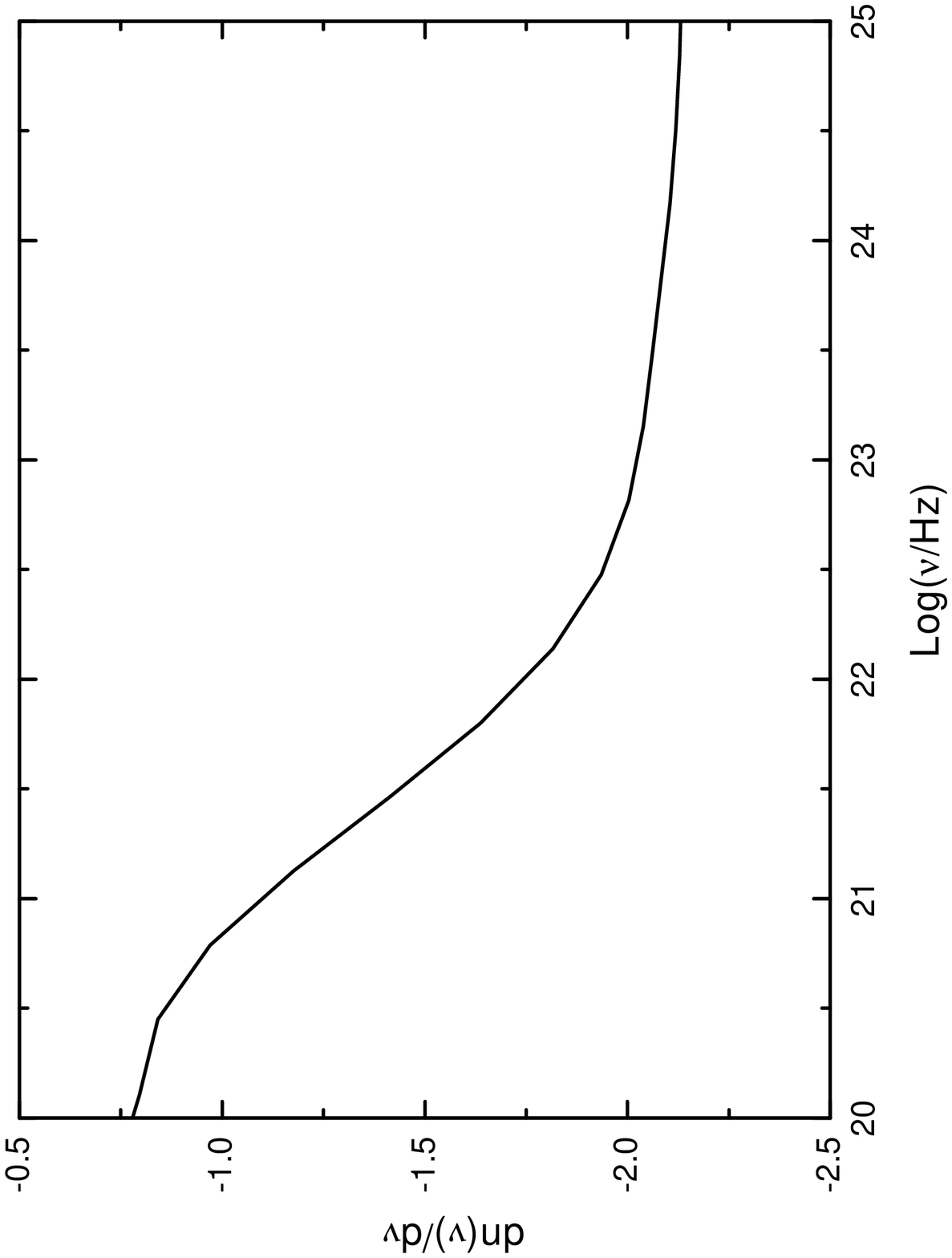}
 \vskip 1.5cm
 \caption{The high energy gamma-ray photon spectrum index $\alpha$ of the SSC
emission from the reverse shock with shock parameters as used in
Fig.1.}
\end{figure}
\noindent

iii) We here compute the slope of the photon spectrum at high
energy bands and plot it in Fig. 3. We can see that at energy
bands from tens of MeV to tens of Gev, the photon spectrum index
$\alpha$ (the photon number $dn(h\nu)/d\nu\propto{\nu^{\alpha}}$)
ranges from 1.7 to 2.15, which is consistent with the observed
high energy gamma-ray photon spectrum by EGRET from some bright
GRBs (e.g. Sommer et al. 1994).

\section{Conclusions and Discussions}
If optical flashes and GeV emissions are, respectively, resulted
from synchrotron and SSC emissions from reverse shocks, they
should show correlations in both their light curves and spectra.
Below we will give the decaying light curve of the synchrotron
self-Compton component of the reverse shock after it has passed
through the ejecta ($t_{obs}>t_{dec}$). Since
$f_m^{rs}\propto{t_{obs}^{-47/48}}\sim{t_{obs}^{-1}}$ and
 $\tau_e^{rs}\propto{t_{obs}^{-1/2}}$,
the peak flux of the inverse Compton spectral component
$f_m^{rs,IC}=f_m^{rs}\tau_e^{rs} \propto{t_{obs}^{-3/2}}$. If the
observed frequency locates between the two break frequencies
($\nu_c^{rs,IC} <\nu<\nu_m^{rs,IC}$), then
$f_\nu^{rs,IC}=f_m^{rs,IC}(\frac{\nu}{\nu_m^{rs,IC}})^{-(p-1)/2}$.
According to Sari \& Piran (1999),
$\gamma_m\propto{t_{obs}^{-13/48}}$ and
$\nu_m^{rs}\propto{t_{obs}^{-73/48}}$, so
$\nu_m^{rs,IC}=2\gamma_m^2{\nu_m^{rs}}
\propto{t_{obs}^{-33/16}}\sim{t_{obs}^{-2}}$, thus
$f_\nu^{rs,IC}\propto{t_{obs}^{-1/2-p}}\propto{t_{obs}^{-3}}$ for
$p=2.5$. On the other hand, if the observed frequency is above
$\nu_c^{rs,IC}$, the flux drops exponentially with time since all
the electrons above the corresponding energy cool and no fresh
electrons are accelerated once the reverse shock has crossed the
ejecta shell. Therefore, in general, we expect to see a rapidly
decaying high energy flux from the reverse shock, which
constitutes a unique characteristic distinguished from other
models  suggested for the observed high energy gamma-ray emission
from some GRBs. Measurements of the time dependence of the high
energy gamma-ray flux  and the spectra with the planned Gamma-ray
Large Area Space Telescope (GLAST) mission will test this
synchrotron self-Compton scenario.

In summary, we showed that SSC process in the reverse shocks of
gamma-ray bursts is a plausible model for the observed prompt
high-energy gamma-ray emissions from several bursts. It is found
that the SSC emission from the reverse shocks dominates over other
emission processes in  energy bands from tens of MeV to tens of
GeV, for a wide range of shock parameters. This model is more
favorable for energetic photons than those related to internal
shocks, since it involves a much lower internal pair-production
optical depth due to a much larger emitting size. We predict that,
in this model, the prompt high-energy emissions are correlated
with the prompt optical flashes, which can be test in the
forthcoming GLAST era

\end{document}